%% file: osned18-sigconf.tex
\documentclass[sigconf]{acmart}

\renewcommand\footnotetextcopyrightpermission[1]{} 
\usepackage{booktabs} 

\usepackage{balance}

\begin{document}


%

\fancyhead{}

\title{Twitter and the Press: an Ego-Centred Analysis}

\author{Chiara Boldrini}
\orcid{1234-5678-9012}
\affiliation{%
  \institution{IIT-CNR}
  \streetaddress{Via G. Moruzzi 1}
  \city{Pisa}
  \country{Italy}
  \postcode{56124}
}
\email{chiara.boldrini@iit.cnr.it}

\author{Mustafa Toprak}
\affiliation{%
  \institution{IIT-CNR}
  \streetaddress{Via G. Moruzzi 1}
  \city{Pisa}
  \country{Italy}
  \postcode{56124}
}
\email{mustafa.toprak@iit.cnr.it}

\author{Marco Conti}
\affiliation{%
  \institution{IIT-CNR}
  \streetaddress{Via G. Moruzzi 1}
  \city{Pisa}
  \country{Italy}
  \postcode{56124}
}
\email{marco.conti@iit.cnr.it}

\author{Andrea Passarella}
\affiliation{%
  \institution{IIT-CNR}
  \streetaddress{Via G. Moruzzi 1}
  \city{Pisa}
  \country{Italy}
  \postcode{56124}
}
\email{andrea.passarella@iit.cnr.it}

\renewcommand{\shortauthors}{C. Boldrini et al.}

\begin{abstract}
Ego networks have proved to be a valuable tool for understanding the relationships that individuals establish with their peers, both in offline and online social networks. Particularly interesting are the cognitive constraints associated with the interactions between the ego and the members of their ego network, whereby individuals cannot maintain meaningful interactions with more than 150 people, on average. In this work, we focus on the ego networks of journalists on Twitter, and we investigate whether they feature the same characteristics observed for other relevant classes of Twitter users, like politicians and generic users. Our findings are that journalists are generally more active and interact with more people than generic users. Their ego network structure is very aligned with reference models derived from the social brain hypothesis and observed in general human ego networks. Remarkably, the similarity is even higher than the one of politicians and generic users ego networks. This may imply a greater cognitive involvement with Twitter than with other social interaction means. Moreover, the ego networks of journalists are much stabler than those of politicians and generic users, and the ego-alter ties are often information-driven.
\end{abstract}

\copyrightyear{2018}
\acmYear{2018} 
\setcopyright{iw3c2w3}
\acmConference[WWW '18 Companion]{The 2018 Web Conference Companion}{April 23--27, 2018}{Lyon, France}
\acmBooktitle{WWW '18 Companion: The 2018 Web Conference Companion, April 23--27, 2018, Lyon, France}
\acmPrice{}
\acmDOI{10.1145/3184558.3191596}
\acmISBN{978-1-4503-5640-4/18/04}

%
%

\ccsdesc[500]{Networks~Online social networks}
\ccsdesc[500]{Human-centered computing~Social network analysis}
\ccsdesc[300]{Mathematics of computing~Graph theory}

\keywords{online social networks, ego networks, Twitter, journalists}

\maketitle

\input{osned18-body}

\bibliographystyle{ACM-Reference-Format}
\balance


\end{document}

%% file: osned18-body.tex
\section{Introduction}
\label{sec:intro}

Online Social Networks are one of the most prominent examples of cyber-physical convergence: social relationships that previously could exist only in the offline world are now transported into the virtual, online dimension and new interactions are enabled that were not possibile before. OSN are thus at the same time very similar and very different from offline social networks, and for this reason several researchers have tried to understand if and to which extent they obey to the same common principles. 

\begin{figure}[ht]
\begin{center}
\includegraphics[scale=0.4]{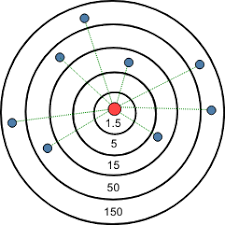}\vspace{-5pt}
\caption{Layered structure of human ego networks}
\label{fig:egonet}\vspace{-10pt}
\end{center}
\end{figure}

Ego networks are the graph-based abstraction that is typically used to study the social relations between an individual and its peers~\cite{everett2005ego,lin2001social,mccarty2002structure,hill2003social}. The ego network is an important abstraction, as it is known that many traits of social behaviour (resource sharing, collaboration, diffusion of information) are chiefly determined by its structural properties~\cite{sutcliffe2012relationships}. In an ego network, the individual, referred to as \emph{ego}, is at the center of the graph, and the edges connect her to the peers (called \emph{alters}) with which she interacts. The ego-alter tie strength is typically computed as a function of the frequency of interactions between the ego and the alter. Grouping these ties by their strength, a layered structure emerges in the ego network (Figure~\ref{fig:egonet}), with the inner circles containing the socially closest peers and the outer circles the more distant relationships. According to the \emph{social brain hypothesis} from evolutionary psycology~\cite{dunbar1998social}, the existence and sizes of the groups are determined by the maximum cognitive capacity of the brain that humans can allocate to \emph{meaningful} social relationships, i.e., beyond the mere level of acquaintances. In this model, five layers exist within the limit of the Dunbar number, which is the maximum number of social relationships (around 150) that an individual, on average, can actively maintain~\cite{hill2003social, Zhou2005}. Beyond the Dunbar number, relationships are just acquaintances and their maintenance has a negligible effect on the cognitive resources. Figure~\ref{fig:egonet} illustrates the typical sizes of each layer (1.5, 5, 15, 50, 150). 

Quite interestingly, ego networks formed through many interaction means, including face-to-face contacts, letters, phone calls, emails, and, remarkably, also OSN, are well aligned with this model~\cite{Zhou2005}. Specifically, Dunbar et al.~\cite{Dunbar2015} have found very similar properties also in Facebook and Twitter ego networks. In this sense, OSN become one of the outlets that is taking up the brain capacity of humans, and thus are subject to the same limitations that have been measured for more traditional social interactions, but are not capable of ``breaking'' the limits imposed by cognitive constraints to our social capacity. Tie strengths and how they determine ego network structures have been the subject of several additional work. For example, in~\cite{gonccalves2011modeling} authors provide one of the first evidences of the existence of an ego network size comparable to the Dunbar's number in Twitter. The relationship between ego network structures and the role of users in Twitter was analysed in~\cite{quercia2012social}. In general, ego network structures are also known to impact significantly on the way information spreads in OSN, and the diversity of information that can be acquired by users~\cite{aral2011diversity}.

Thanks to its public API and its impressive number of users (330 million active users as of October 2017\footnote{Source: Wikipedia}), Twitter is one of the most studied OSN in the related literature on social networking. Generic users on Twitter have received a lot of attention, where the aim was to characterise how they engage with the platform~\cite{java2007twitter:-understanding}, interact with each other, and how real life affects what we observe on Twitter and vice versa~\cite{tumasjan2010predicting}. However, users on Twitter are not created equal: their behaviour on the platform is intertwined to their social status and their job. For this reason, researchers have recently focused their attention to special classes of Twitter users that, for one reason or another, can be singled out. For example, Arnaboldi et al.~\cite{Arnaboldi2017} have studied how politicians engage with other people on Twitter, highlighting that, while the same ego network structure appears, there is a significant difference with respect to generic users in terms of ego network variability and amount of information-driven relationships.

Another interesting category of Twitter users is that of journalists. Previous research has highlighted how journalists use Twitter as a platform for personal branding~\cite{canter2015personalised-tweeting} and to promote content from their news websites~\cite{russell2015sets}. In this sense, journalists share with politicians many of the same motivations for using Twitter. However, no previous work has studied how journalists create and maintain social relationships on Twitter, and whether this resembles what we observe for similar categories of users or for generic users. Thus, in this paper, exploiting the ego network abstraction described above, we investigate the activity of journalists on Twitter in terms of how they engage with other users. 

We found that journalists are significantly more active than the average generic users and that they have a marked preference for social communications involving retweets. From the ego network's standpoint, journalists feature a layered structure that is is very similar to the general ego network model of Figure~\ref{fig:egonet}, sometimes much more than what has been observed for other types of Twitter users. Specifically, the size of the active network is much closer to the theoretical value of 150, and very similar to the one observed in offline ego networks before the diffusion of OSN. This is a surprising result, that may indicate that journalists invest a large portion, compared to other types of users, of their social capacity on Twitter. The ego networks of journalists are also very stable over time, and this is very different from the findings about politicians and generic users on Twitter~\cite{Arnaboldi2017}. This may be due to the fact that journalists use Twitter less as a leverage to win people over and more as a social communication tool. Finally, we show that many ego-alter relationships are information-driven:  they are often activated by a hashtag, their future interactions also involve hashtags, and this holds true approximately for all the layers with the same intensity.

\section{The dataset}
\label{sec:dataset}

We downloaded the timeline for the journalists belonging to the Italian journalists list at \url{https://twitter.com/stampa_tweet/lists/giornalisti}. This list comprises 492 journalists that are the most popular in Italy. Due to the Twitter API limitations, only the 3200 most recent tweets could be downloaded for each journalist. Of these 492 members, we discard those with a protected timeline and with zero tweets, ending up with 486 journalists. All of them are long-time Twitter users, since their most recent registration dates back to 2012. However, as we will discuss later on, not all of them are equally engaged with the Twitter platform.

\subsection{Observability}

The 3200 tweets limitation imposed by the Twitter API affects journalists' timelines non uniformly. Specifically, around $69\%$ of users (corresponding to 335 out of the 486 journalists) have posted more tweets than what we are able to download. We refer to these users as \emph{partially observed}. This is a significant difference with respect to the dataset analysed in~\cite{Arnaboldi2013}, where, for $98\%$ of users, 3200 tweets were enough to cover the whole Twitter activity. Please note that being partially or fully observed is an indirect measure of the user's tweeting frequency: the more the tweets posted, the quicker the 3200 slots are saturated. Indeed, the average daily tweeting frequency is 0.6 for the fully observed users and 6.0 for the partially observed ones (which, anyway, is still a feasible value for truly human Twitter activity).

\begin{figure}[ht]
\begin{center}
\includegraphics[scale=0.5]{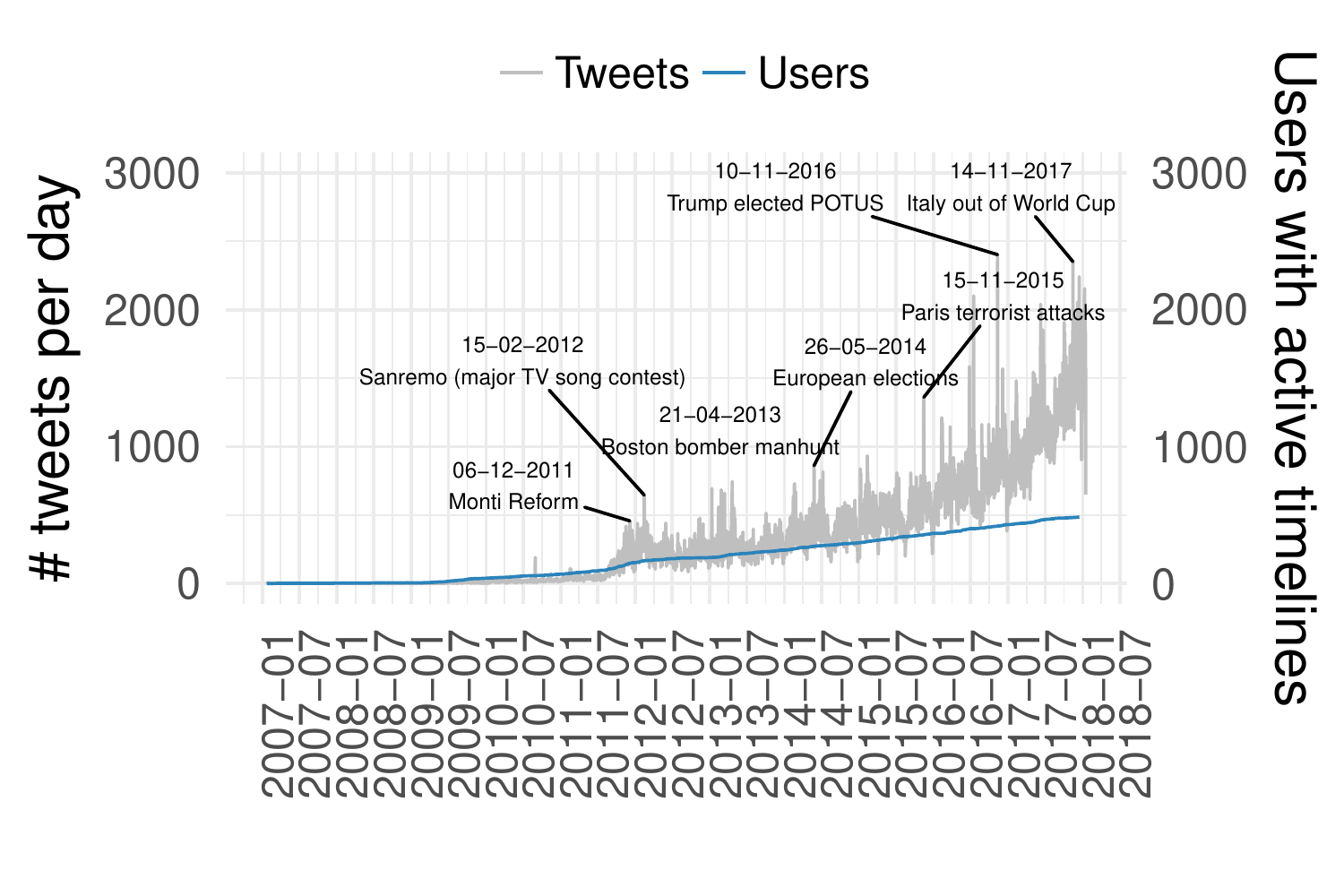}\vspace{-10pt}
\caption{Tweeting volume VS active users over time}
\label{fig:tweet_volume_evolution}\vspace{-5pt}
\end{center}
\end{figure}

We refer to the portion of timeline that we are able to observe as \emph{observed timeline}. When the 3200 tweets limitation does not kick in, the observed timeline overlaps with the \emph{active timeline}, which we estimate as the Twitter activity since the user registration on Twitter\footnote{The real active timeline would be the time between the first and last tweet. However, in this case the time of the first tweet is unknown due to the API limitations.}. While the distinction between partially observed and fully observed users is binary, the actual coverage of the observed timeline with respect to the active timeline is more nuanced. Hence, the percentage of active timeline observed typically changes on a per user basis. 
%
%
%
%
The direct consequence of the uneven time coverage of the observed timelines is that the number of tweeting journalists in our dataset constantly increases over time starting from 2007, because the most active users are present only in the most recent intervals. Figure~\ref{fig:tweet_volume_evolution} shows how this impacts on the daily volume of tweets generated over time. We plot in gray the time series of total observed tweets per day, while in blue we plot the number of users with active observed timeline (i.e., for which the coverage of the 3200 tweets has already started). While the growth in the the number of active users is approximately linear, the increase in the observed tweets is exponential. This further highlights one of the characteristics of this dataset: the most active users are only observed for a short amount of time but they are the ones that impact the most on the number of generated tweets.

\subsection{Data cleaning and filtering}
\label{sec:data_cleaning}

Not all the downloaded timelines are suitable for either a general analysis of the behaviour of journalists on Twitter or for the ego network analysis.  In fact, some journalists may have left Twitter long ago, others may use it only sporadically, others may feature an unstable ego network (we will discuss what this means in a later section). In the following, we discuss how to identify the Twitter users that should be considered for our analysis.

\subsubsection{Users that have left Twitter}
\label{sec:twitter_abandonment}

First, we focus on identifying users that are not actively engaged with the platform anymore. 
The \emph{inactive life} of a user, defined as the time period between the last observed tweet and the time of downloading, is an indirect measure of inactivity. In fact, long inactive periods are a telltale sign of decreased engagement. 

In previous works~\cite{Arnaboldi2013}, a threshold-based approach mimicking Twitter's 6-months inactivity criterion is employed: a user was considered to have abandoned Twitter if her inactive life was longer than six months. However, here we argue that six months of inactivity may be perfectly in line with the historic behaviour of a user, without being suggestive of her having abandoned the platform. Therefore, we use a new classifier based on the concept of \emph{intertweet time}, defined as the time interval between two consecutive tweets. For each journalist $i$, we observe a certain characteristic distribution of intertweet times ($ITT_i$). If the inactive lifespan $T^{inactive}_i$ of user $i$ is significantly larger than the maximum intertweet time previously observed for the user, then it is likely that the user has indeed abandoned the platform. Specifically, we allow for a a six-months grace period (in the spirit of the considerations in~\cite{Arnaboldi2013}) and we assume a user $i$ is still active if $T^{inactive}_i < \max\{ITT_i\} + \textrm{6 months}$. Using this classifier, 11 users are marked as having abandoned Twitter at the time we downloaded the dataset.

\subsubsection{Regular users}
\label{sec:active_users}

\begin{figure}[ht]
\begin{center}
\includegraphics[scale=0.5]{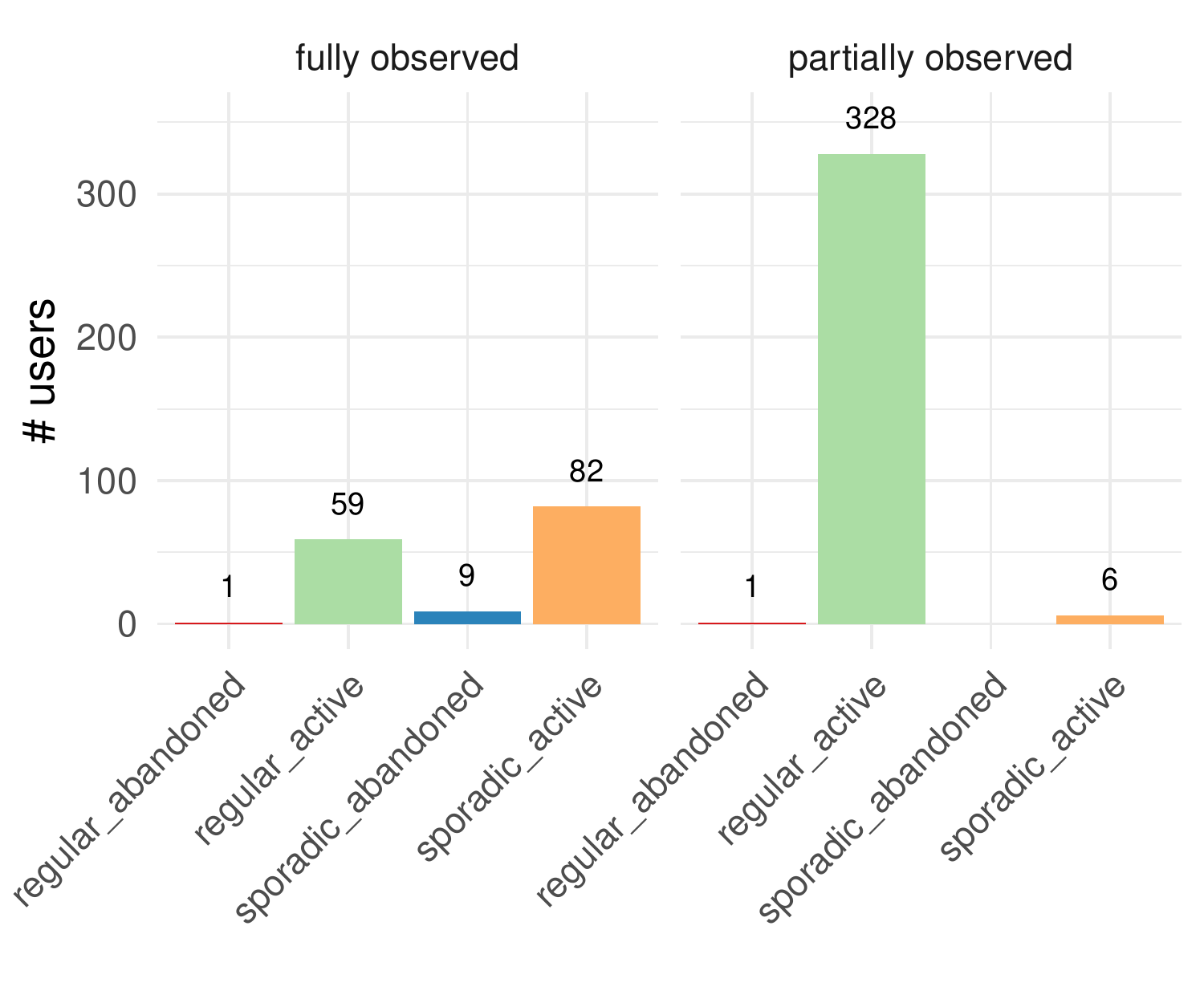}\vspace{-15pt}
\caption{User classification}
\label{fig:user_classifier}\vspace{-10pt}
\end{center}
\end{figure}

While the previous classifier was concentrating on the inactive life of a user, we now focus on the observed life, and we investigate the level of regularity in  Twitter activity. Like in~\cite{Arnaboldi2017}, we assume that a Twitter user is a regular user if she posts at least one tweet every 3 days (as an average frequency within each month) for at least $50\%$ of the total number of months of their activity. If she does not, the user is classified as sporadic. In Figure~\ref{fig:user_classifier} we combine the results of the abandonment classifier and the regularity classifier. As expected, partially observed users are generally more active and regular than fully observed ones. It is interesting to note that there are two users marked as regulars that have then stopped using Twitter. We manually checked these two users to confirm the classification and we found out that indeed one of them had died and one had changed its Twitter handle. Occasional users are not interesting for our analysis, since their engagement with the platform is low and so is their cognitive involvement. For this reason, in the following we will focus on regular and active users only (corresponding to the green bar in Figure~\ref{fig:user_classifier}), which amount to 387 journalists in total. 

\subsubsection{Stationarity}
\label{sec:stationarity}

\begin{figure}[ht]
\begin{center}
\includegraphics[scale=0.45]{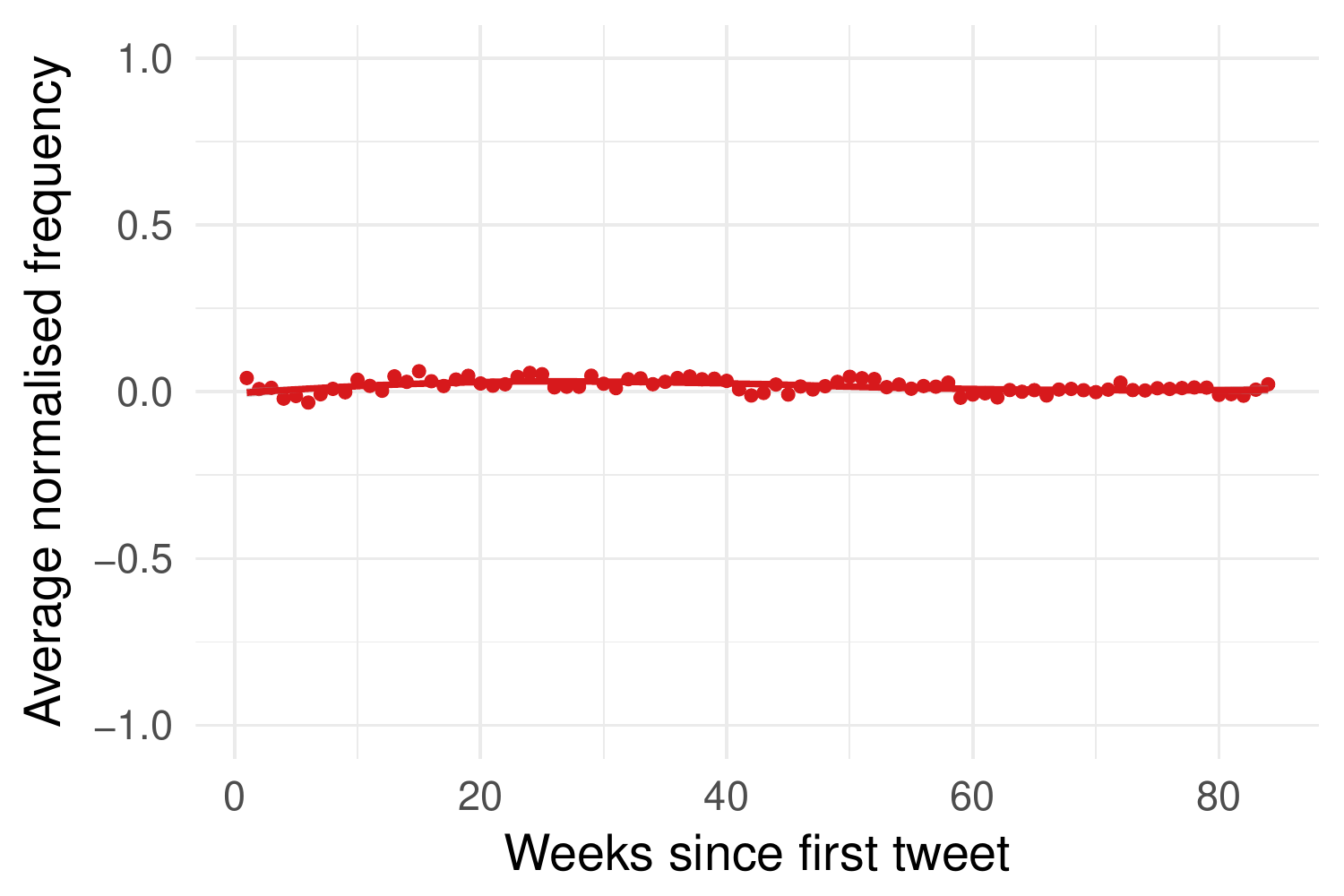}\vspace{-5pt}
\caption{Stationarity analysis}
\label{fig:stability}\vspace{-10pt}
\end{center}
\end{figure}

In our analysis, we are not only interested in active and regular users, but we also need to make sure that their Twitter activity has already stabilised. In fact, it is a frequent finding in the related literature that users tend to be more active during the initial interactions with the platform, reaching a steady state later on, when the engagement is somewhat consolidated. In order to verify whether this is the case also in our dataset, we compute for each user the average number of tweets generated each week. For this analysis, we align all observed timelines to start at the same time (i.e., time 0 for each given user is the time of the first observed tweet of the user). Then, we average these values across all users. Please note that we restrict our analysis to the first 84 weeks, since the number of active users abruptly decreases afterwards. For the sake of a fair comparison between users, we have rescaled the weekly frequencies of each user using the mean normalization technique. This procedure yields values in the range $[-1,1]$, with values around zero being close to the average frequency. The results are shown in Figure~\ref{fig:stability}, only for the 59 active and regular users that are fully observed\footnote{Please note that for partially observed users we are missing the initial portion of the timeline, so this analysis cannot be replicated.}.  It is clear that there is not a significant variation in the activity between first and last portion of the timelines. Hence, we do not discard a transient period for the users in our dataset.

\subsection{Dataset overview}

\begin{figure}[ht]
\begin{center}
\includegraphics[scale=0.5]{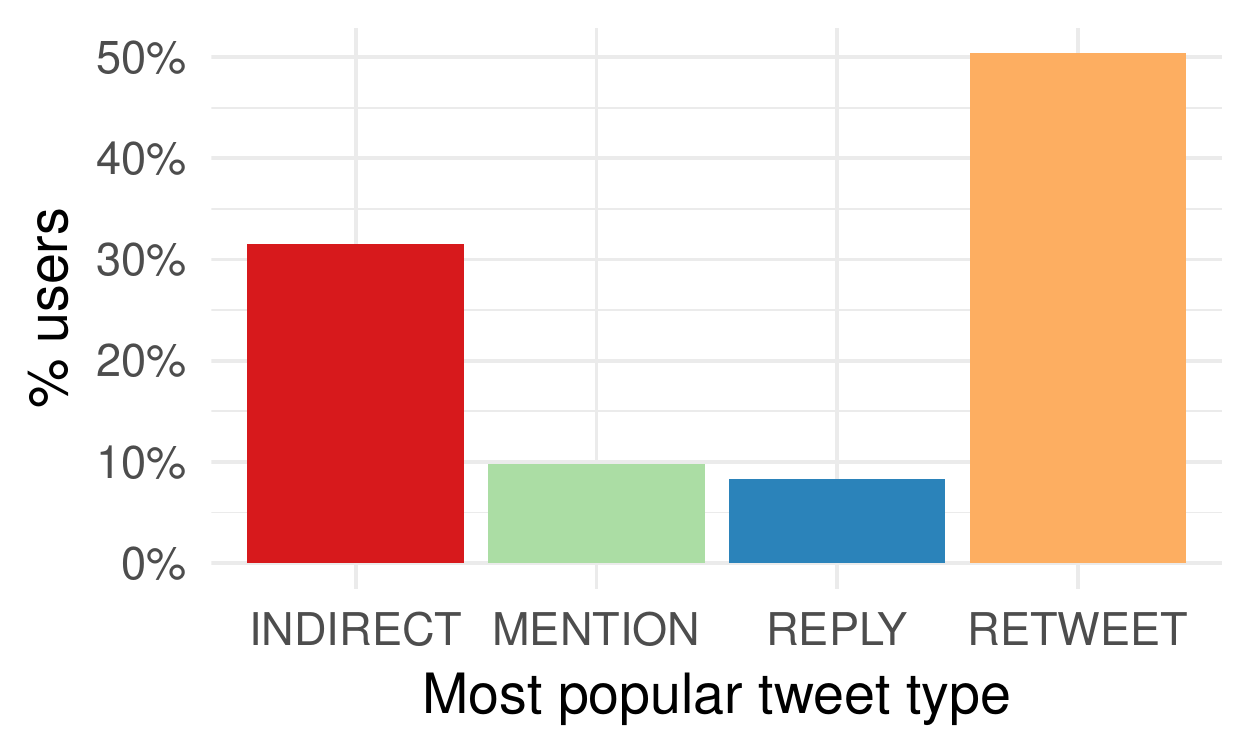}\vspace{-5pt}
\caption{Most popular tweeting activity}\vspace{-10pt}
\label{fig:top_activity}
\end{center}
\end{figure}

\begin{figure}[ht]
\begin{center}
\includegraphics[scale=0.45]{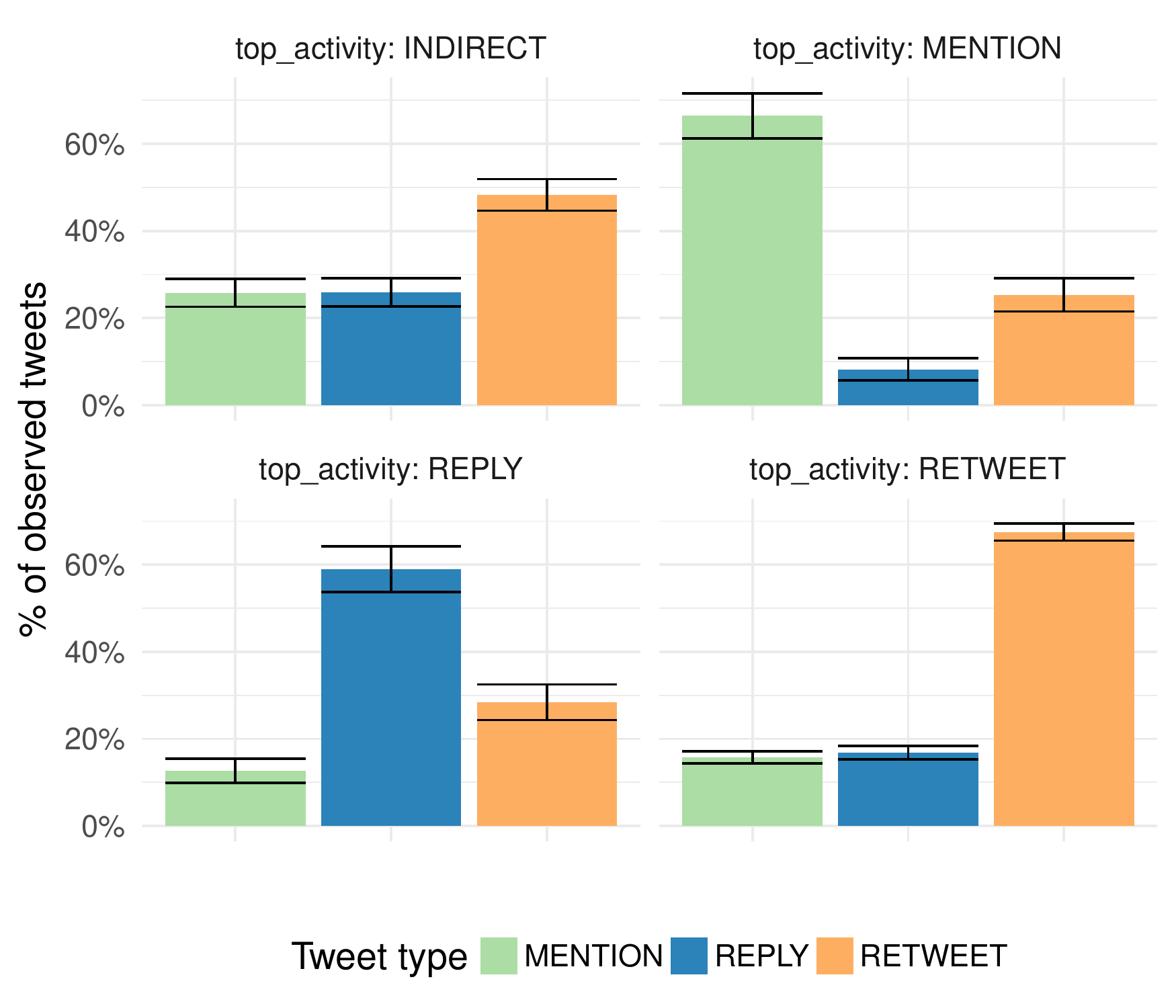}\vspace{-5pt}
\caption{Tweets per top activity}
\label{fig:types_per_top_activity}\vspace{-15pt}
\end{center}
\end{figure}

The main properties of the dataset resulting from the filtering performed in the previous section are summarised in Table~\ref{tab:summary}. With an average active life of about 8 years, all these journalists are long-time Twitter users. However, the number of tweets that they have generated during their entire life varies significantly between users. Among the tweets that we observe, an average of $68\%$ are social, i.e., following the definition in~\cite{Arnaboldi2013}, involve forms of direct communications such as replies, mentions or retweets. Among these, retweets\footnote{Please note that, for the purpose of this analysis, quote tweets are treated as simple retweets.} are by far the most popular mode for interaction, taking up half of the social communication space. Retweets were popular also for the politicians analysed in~\cite{Arnaboldi2017}, but their percentage was around $40\%$, with mentions being used, by and large, with roughly the same frequency.  The popularity of retweets is confirmed when looking at the most popular communication mode per user (Figure~\ref{fig:top_activity}, where \emph{indirect} means tweets that are neither mentions, nor replies, nor retweets): retweets are the go-to mode for more than $50\%$ of the users. Figure~\ref{fig:types_per_top_activity} shows the breakdown across the three types of social tweets, grouping users by their dominant type of tweet. It shows that retweets rank always as first or second most popular social mode, regardless of the predominant mode.

\begin{table*}[t]
\centering
\caption{Summary statistics}
\begin{tabular}{rrrrrrrrr}
  \hline
 & active life [years] & tot tweets & observed tweets & tweets/day & $\%$ social  & $\%$ replies & $\%$ retweets & $\%$ mentions \\ 
  \hline
mean & 7.88 & 16,319 & 3061.34 & 5.33 & 68.26 & 22.33 & 54.06 & 23.62 \\ 
sd & 1.33 & 22,199 & 435.06 & 9.27 & 21.11 & 18.35 & 22.02 & 20.03 \\ 
   \hline
\end{tabular}
\label{tab:summary}
\end{table*}

\begin{figure}[ht]
\begin{center}
\includegraphics[scale=0.5]{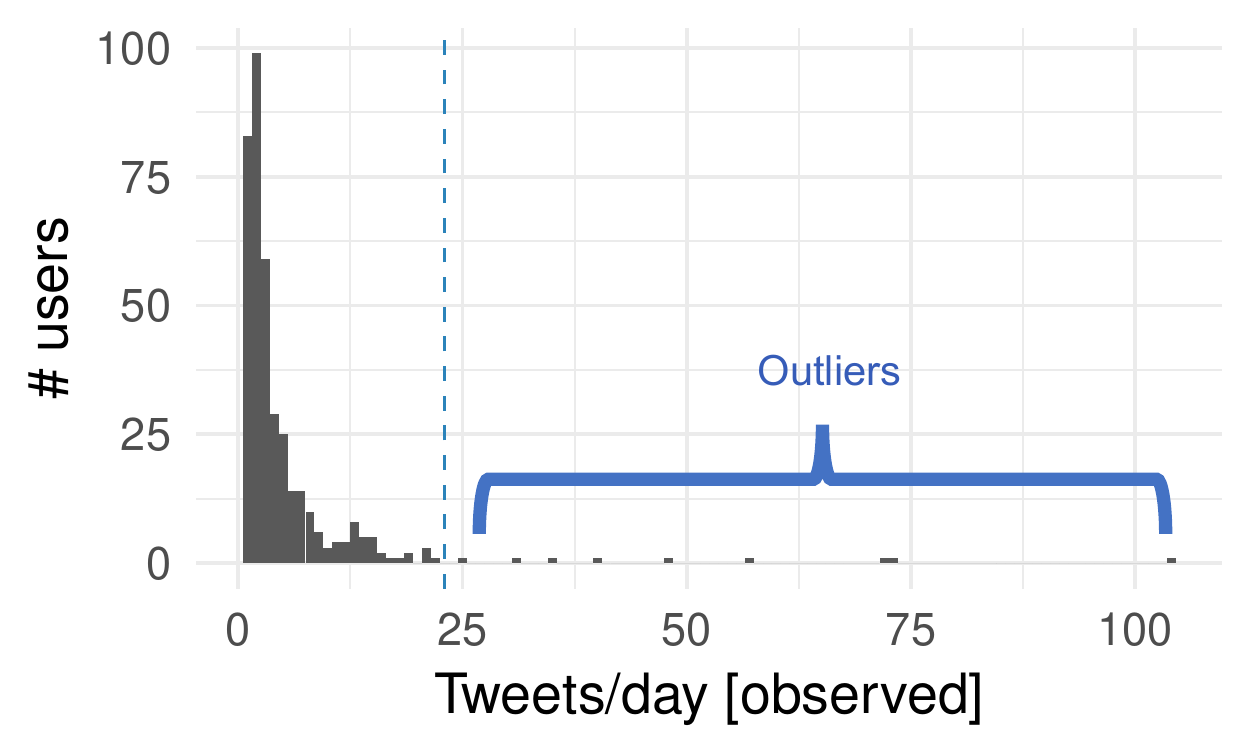}\vspace{-5pt}
\caption{Histogram of tweet frequency}\vspace{-10pt}
\label{fig:twitter_activity}
\end{center}
\end{figure}

\begin{figure}[ht]
\begin{center}
\includegraphics[scale=0.5]{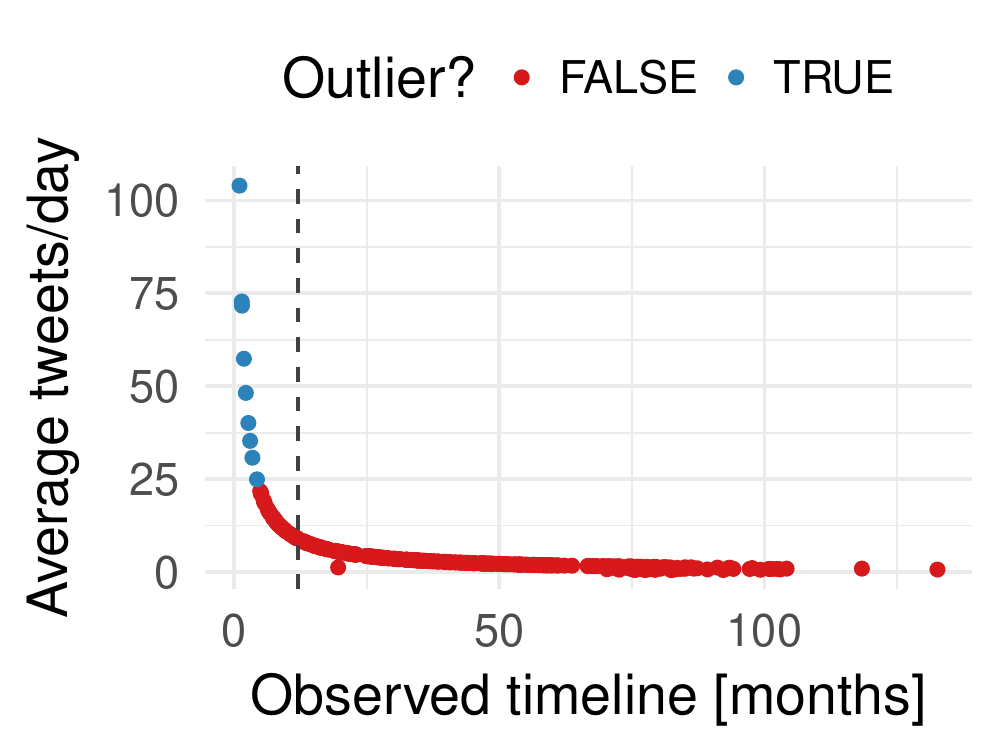}\vspace{-5pt}
\caption{Observed timeline length VS average tweet frequency}
\label{fig:scatter_observedtl_vs_freq}\vspace{-10pt}
\end{center}
\end{figure}

Table~\ref{tab:summary} also reveals a great variability in the number of tweets per day, whose distribution is shown in Figure~\ref{fig:twitter_activity}. Figure~\ref{fig:twitter_activity} shows that while most journalists post around 10-15 tweets per day, some users are clear outliers, with daily tweets well above 30. These users are marked as outliers when using standard clustering techniques such as DBSCAN on the tweet frequency. These outliers could have a significant impact on the ego network statistics, due to their intense Twitter activity, and should be studied separately. However, as shown in Figure~\ref{fig:scatter_observedtl_vs_freq}, the observed timeline for these very active users is too short to satisfy the stability requirement (at least one year between the first and last observed tweets) for ego networks that has been typically used in the related literature~\cite{Arnaboldi2013}. For this reason, these users will be excluded from the analysis in Section~\ref{sec:egonets_analysis}. The average tweet frequency when excluding outliers is 4.17 tweets per day, which is still much higher than the one observed\footnote{Please note that the comparison between our results about journalists and those in~\cite{Arnaboldi2017} about politicians is a fair one. In fact, the filtering based on user regularity and activity intensity described in Section~\ref{sec:data_cleaning} was also performed very similarly in~\cite{Arnaboldi2017}. More specifically, the filtering described in Section~\ref{sec:active_users} is exactly the same as the one used in~\cite{Arnaboldi2017}. Our filter described in Section~\ref{sec:twitter_abandonment} is slightly more conservative than the one used in~\cite{Arnaboldi2017}, hence it potentially retains more users that are less active. Despite this, as discussed later in the paper, we find that journalists are generally more active than politicians and generic users.} in politicians (between 2.6 and 3.2 depending on the dataset~\cite{Arnaboldi2017}).

\section{Ego network analysis}
\label{sec:egonets_analysis}

We now focus our attention on the ego networks of the journalists that have been selected according to the discussion in Section~\ref{sec:dataset}. For each of these journalists, the strength of ego-alter ties are inferred from the frequency of \emph{direct tweets} (mentions, reply, or retweets) between the ego and the alters, which is normally considered as a proxy for their emotional closeness~\cite{Arnaboldi2013,Dunbar2015}. 
Similarly to the related literature, we define as active tie one for which the ego and the alter exchange at least one message per year.

\subsection{Static ego networks}
\label{sec:egonets_analysis_static}

The static view of an ego network is a single, aggregate snapshot of the ego and its ties. By modelling the ego-alter ties through a static ego network, we consider all the communications between the ego and the alter. This analysis provides an essential starting point for understanding the ego-alter interactions and for making quantitative comparison with other datasets studied in the related literature.

\begin{figure}[ht]
\begin{center}
\includegraphics[scale=0.45]{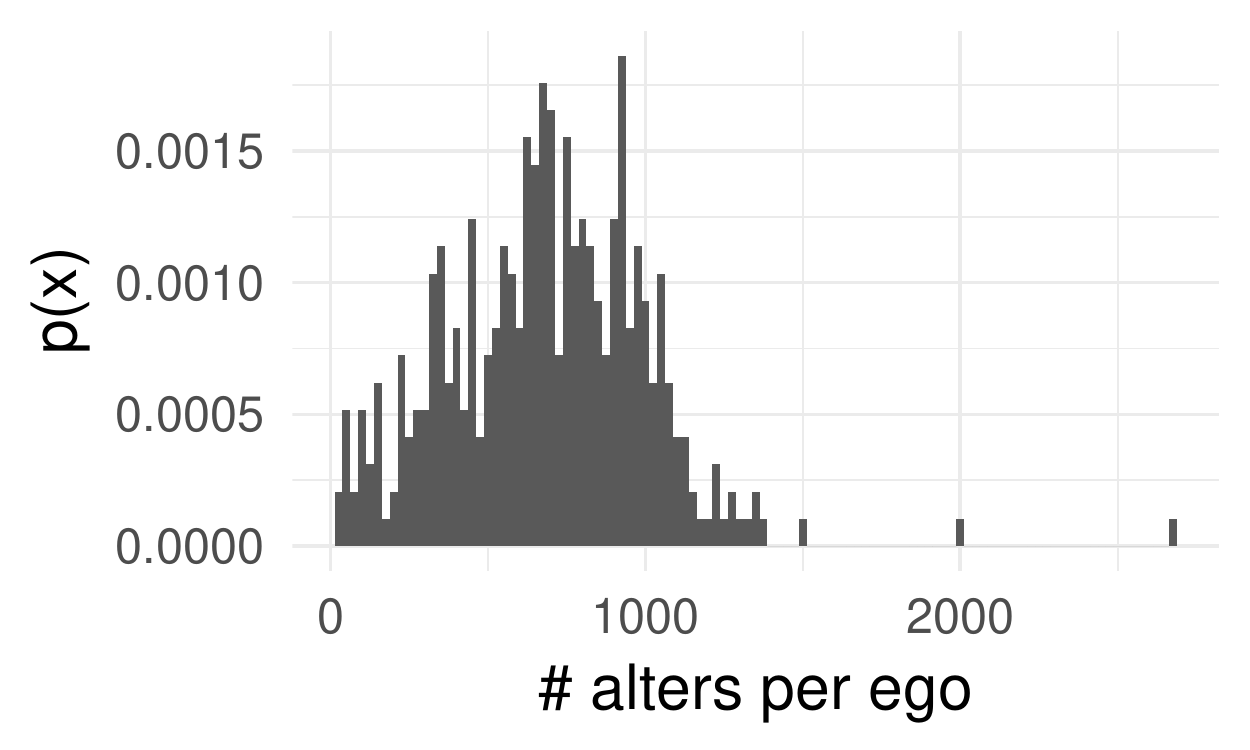}\vspace{-10pt}
\caption{Histogram of number of alters per ego}
\label{fig:total_size}\vspace{-10pt}
\end{center}
\end{figure}

\begin{figure}[ht]
\begin{center}
\includegraphics[scale=0.45]{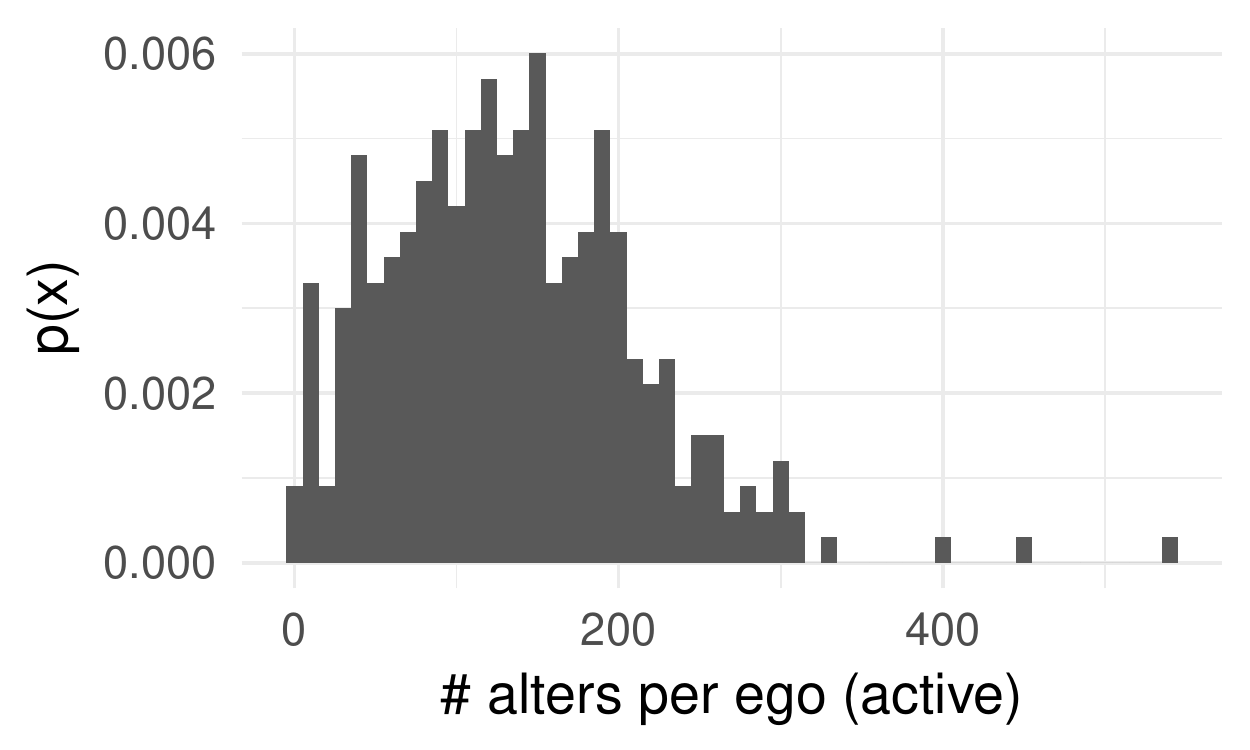}\vspace{-10pt}
\caption{Histogram of active networks size}\vspace{-10pt}
\label{fig:activenet_size}
\end{center}
\end{figure}

Figure~\ref{fig:total_size} shows the distribution of the number of alters per ego, i.e., the total number of social ties, including both active and inactive ones. When compared with what was observed in ~\cite{Dunbar2015} for generic Twitter users, we notice that, for journalists, the number of alters per ego is about twice as large as that of regular users. This implies that journalists tend to contact significantly more alters. This is in line with journalists being prominent users that stand out among the Twitter crowd. However, as discussed before, only active relationships consume cognitive resources and thus are affected by the human cognitive constraints. For this reason, in Figure~\ref{fig:activenet_size}, we consider the active network, i.e., the ego network encompassing only relationships with contact frequency higher than one tweet per year. While the size of the active network is substantially different from that of the complete ego network, the proportional difference is maintained when comparing what we observe for journalists against the results shown in~\cite{Dunbar2015} for generic Twitter users\footnote{While the initial filtering used in~\cite{Dunbar2015} is different from the one discussed in this paper (Section~\ref{sec:data_cleaning}), the comparison between the two sets of results is still meaningful. In fact, the filtering in~\cite{Dunbar2015} (whereby only users with an average of more than 10 interactions per month are kept) tends to retain only very active users, thus it might overestimate the social interactions of generic users. However, our results indicate that journalists tend to be even more engaged than this active subset of generic users.}.

\begin{figure}[ht]
\begin{center}
\includegraphics[scale=0.45]{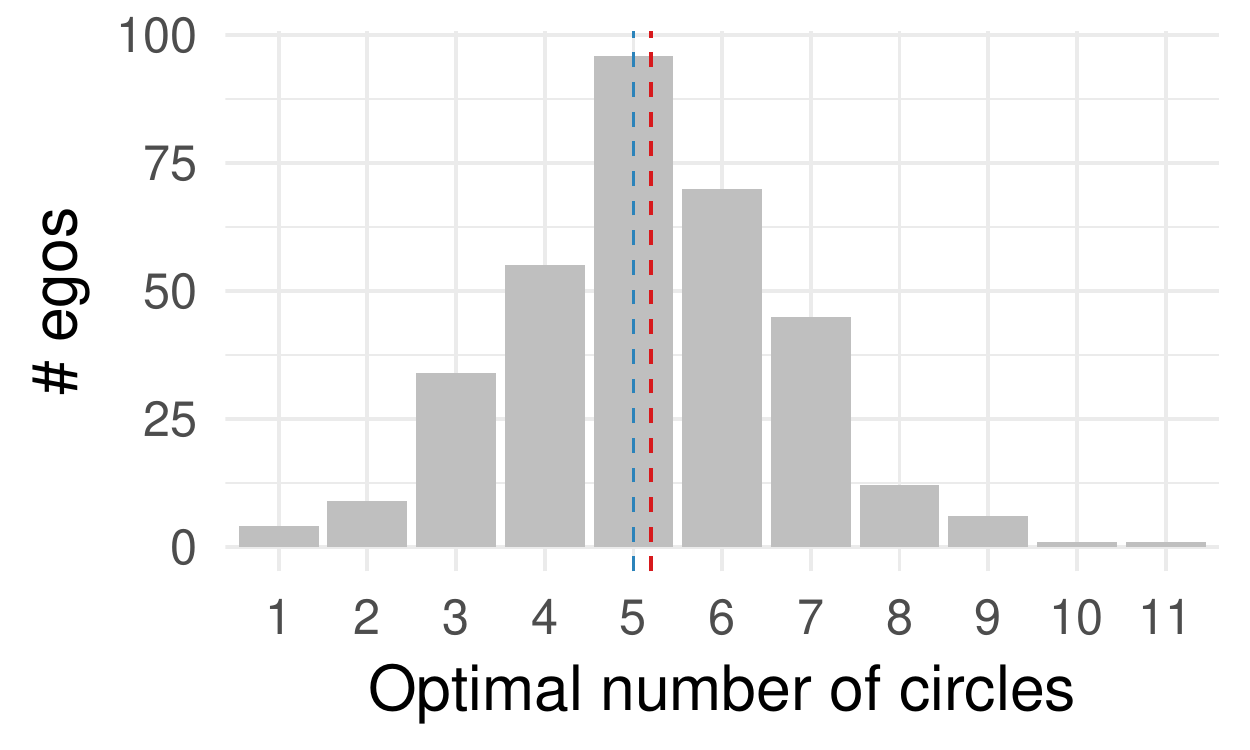}\vspace{-10pt}
\caption{Optimal number of circles per ego (red line: average, blue line: median)}
\label{fig:optimal_circles}\vspace{-10pt}
\end{center}
\end{figure}

Each of these active networks can be partitioned into a set of layers (also called \emph{circles}), grouping the tie strength between the ego and its alters. Offline social networks typically feature four layers~\cite{Zhou2005}, while five layers are a common finding for OSN~\cite{Dunbar2015}. The optimal number of layers can be obtained through cluster analysis. Specifically, here we use the Mean Shift algorithm~\cite{comaniciu2002mean}. As shown in Figure~\ref{fig:optimal_circles}, similarly to other OSN, the optimal number of circles for the egos in our datasets is five. 

\begin{figure}[ht]
\begin{center}
\includegraphics[scale=0.45]{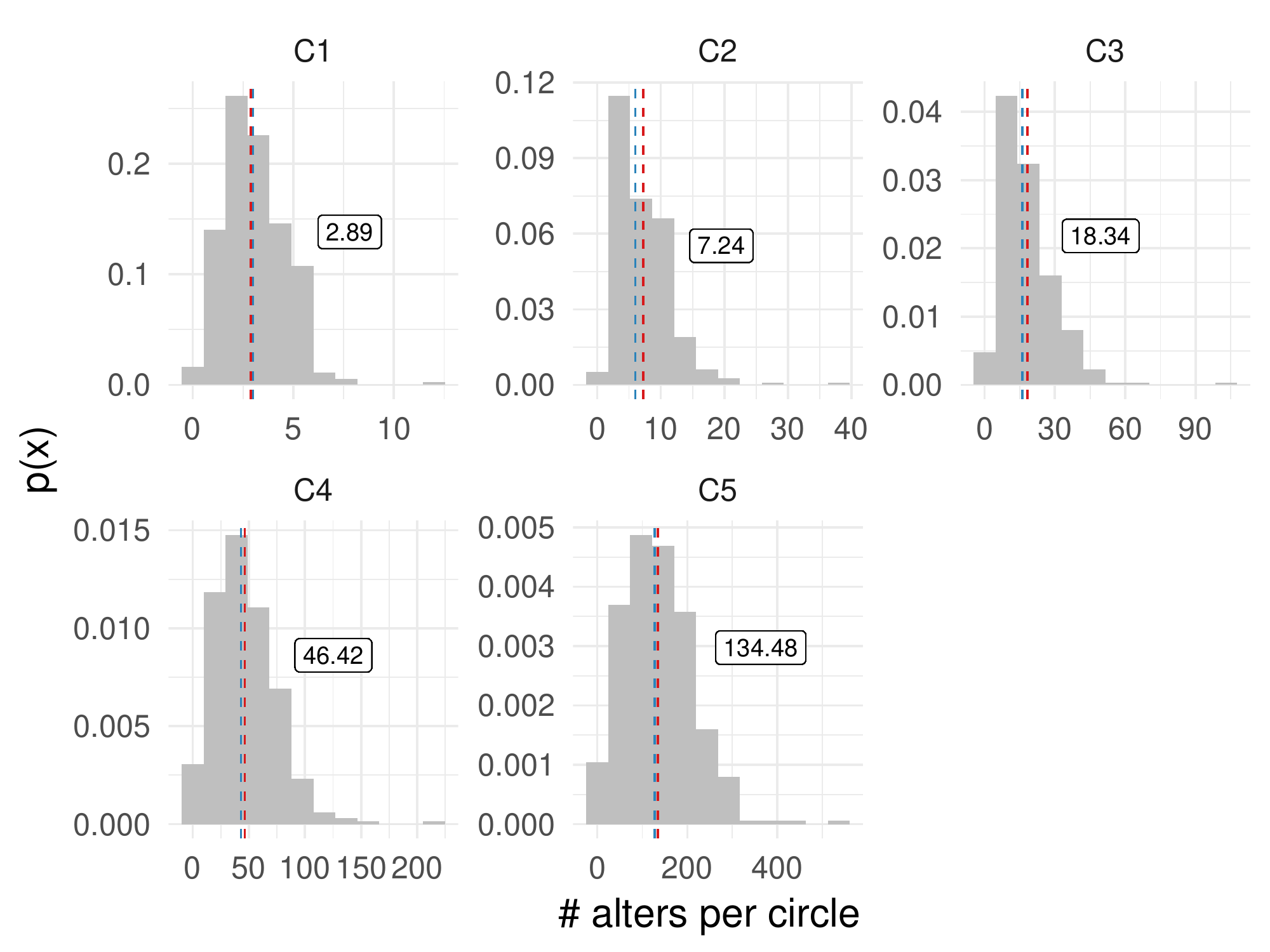}\vspace{-10pt}
\caption{Distribution of circle sizes}
\label{fig:circles_size}\vspace{-10pt}
\end{center}
\end{figure}

When looking at the composition of circles (Figure~\ref{fig:circles_size}), we obtain average values that are very close to the sizes of social groups (1.5, 5, 15, 50, and 150) predicted by Dunbar for offline social networks. While the layer sizes for OSN were found to be typically smaller than those in offline social networks~\cite{Dunbar2015}, the journalists group seems instead to mimic very closely the offline dynamics. Specifically, the sizes of the two layers are slightly larger than in the model. Note, however, that this might depend on the specific clustering algorithm used. Similar deviations across clustering techniques have been observed also before~\cite{Dunbar2015}. On the other hand, the sizes of the two outermost layers are significantly larger for journalists than for generic users~\cite{Dunbar2015} and politicians~\cite{Arnaboldi2017}. Remarkably, the average size of the entire ego network is very close to the first empirical observation ($\sim130$) that validated the social brain hypothesis and provided evidence for the reference size of 150 alters~\cite{hill2003social}.

\begin{figure}[ht]
\begin{center}
\includegraphics[scale=0.45]{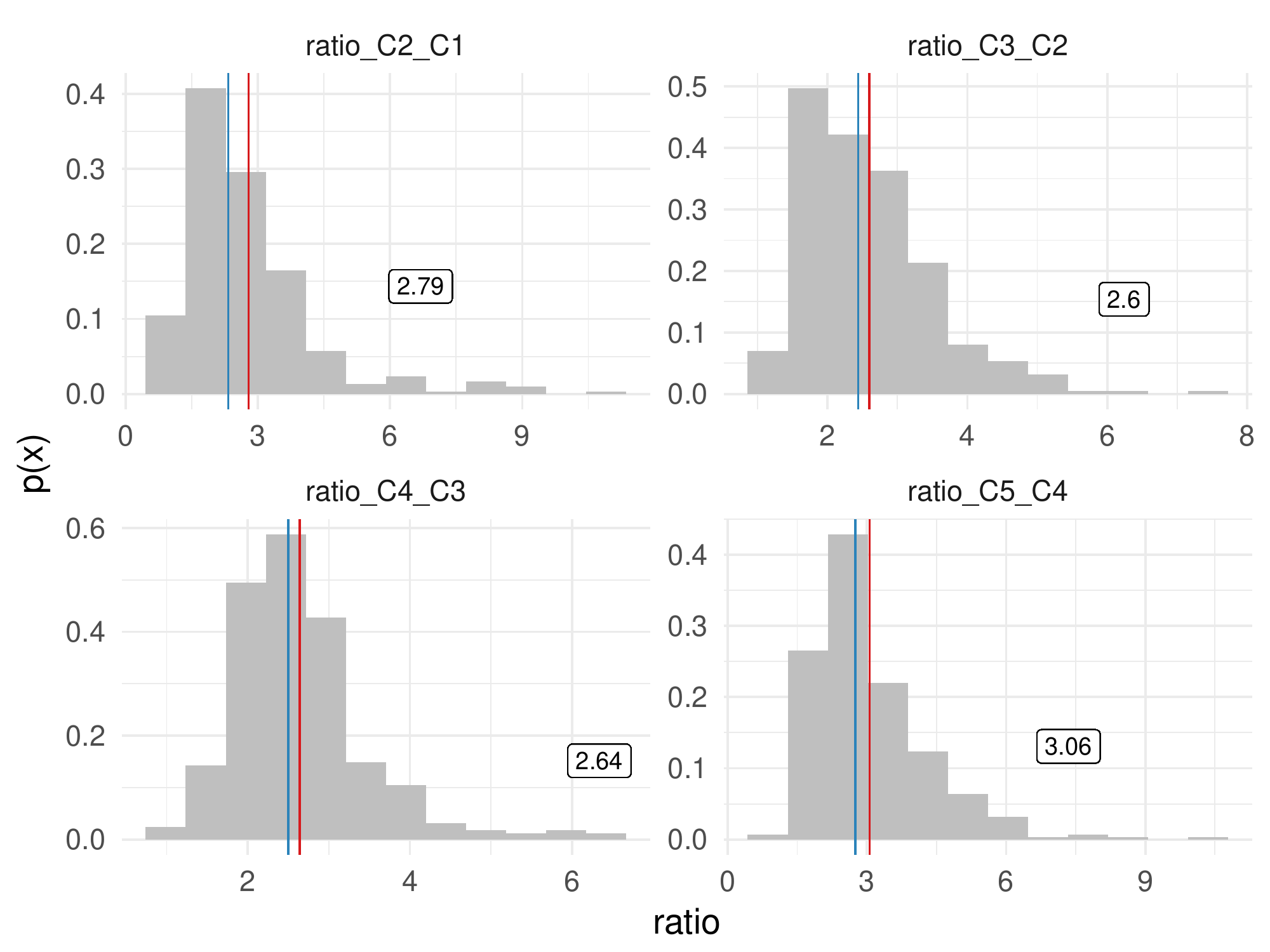}\vspace{-10pt}
\caption{Distribution of scaling ratios}
\label{fig:scaling_ratios}\vspace{-10pt}
\end{center}
\end{figure}

Figure~\ref{fig:scaling_ratios} shows the scaling ratios between adjacent circles, confirming the remarkable match with the characteristic value of 3 commonly found in the related literature.

\subsection{Dynamic ego networks}
\label{sec:egonets_analysis_dynamic}

The dynamic analysis of ego networks focuses on their evolution over time. Specifically, snapshots of one year are considered \footnote{Hence, egos for which we observe less than two years of tweets are filtered out}, each being shifted forward by one month with respect to the previous one. For each snapshot we focus on the alters that are members of each ring, where a ring is defined as the portion of circle that excludes its inners circles. 

The overlapping between the members of the same ring across different snapshots is measured using the Jaccard index. The Jaccard index is obtained dividing the cardinality of the intersection set  by the cardinality of the union set. The closer this index to $1$, the better the overlapping. The amount of movements between rings is measured through the Jump index, which simply counts (and then averages across all alters in the ring) the number of jumps between rings. 

\begin{figure}[ht]
\begin{center}
\includegraphics[scale=0.45]{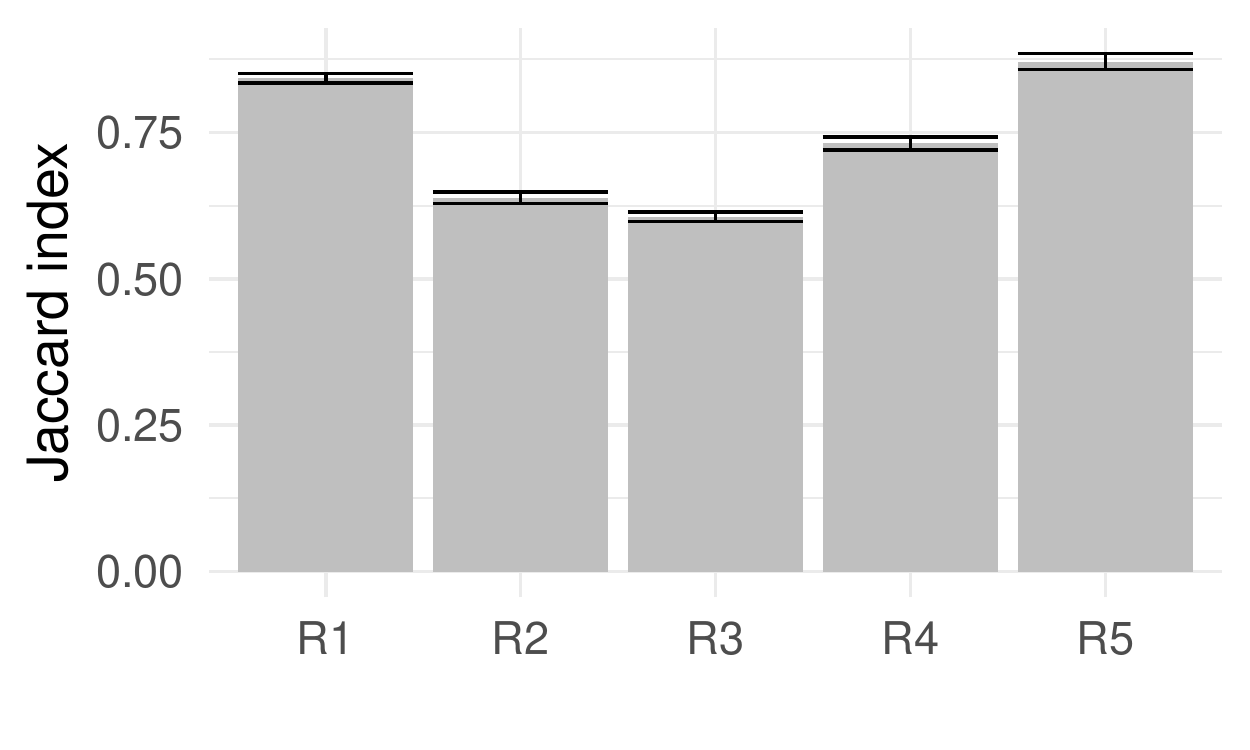}\vspace{-15pt}
\caption{Average Jaccard index per ring}\vspace{-10pt}
\label{fig:jaccard}
\end{center}
\end{figure}

\begin{figure}[ht]
\begin{center}
\includegraphics[scale=0.45]{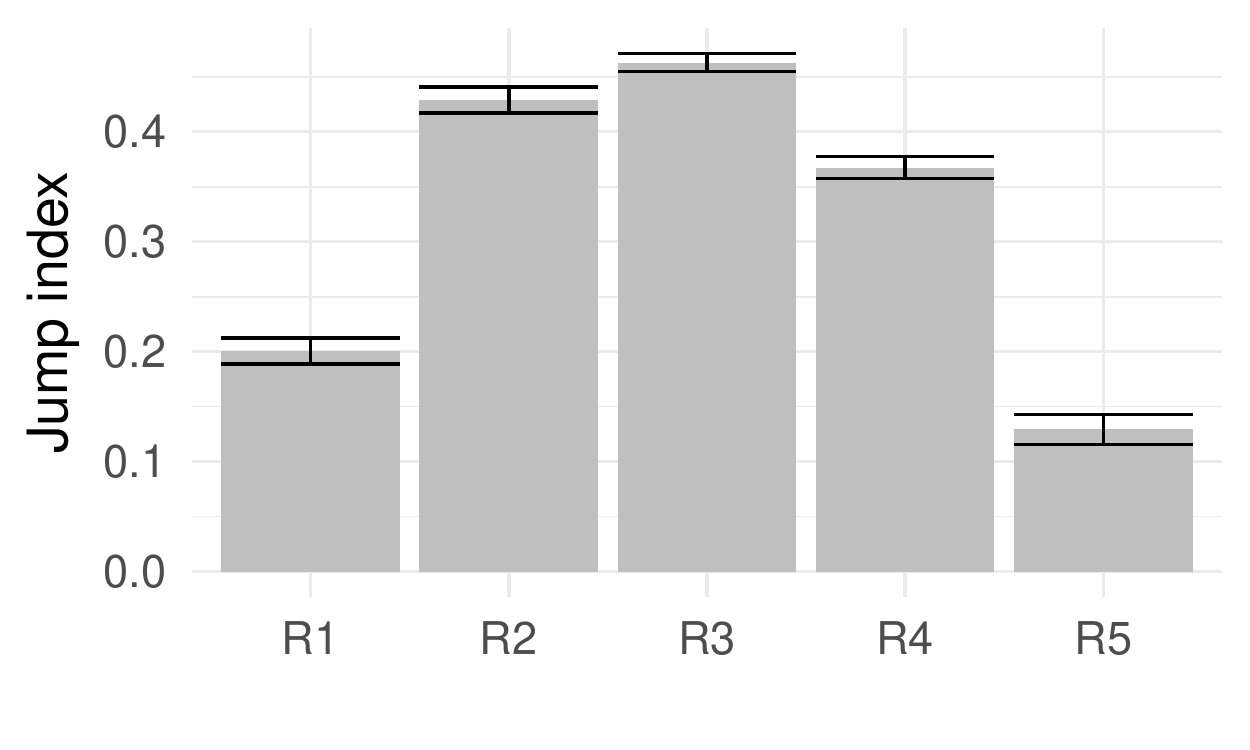}\vspace{-15pt}
\caption{Average jump index per ring}\vspace{-10pt}
\label{fig:jump}
\end{center}
\end{figure}

Figures~\ref{fig:jaccard} and~\ref{fig:jump} show the values of these indices in each of the five rings. They reveal very stable layers, in which the composition does not change much and in which jumps between rings are very few. This is in stark contrast with the results obtained for the politicians in~\cite{Arnaboldi2017}. In fact, politicians features low stability in all rings (but especially in the inner most ones), and their jump index is high. We conjecture that journalists may use Twitter less as a leverage to win people over and more as a social communication tool. Further analyses are required to substantiate this claim.

\subsection{Social tweets and hashtags}
\label{sec:hashtags}

\begin{figure}[ht]
\begin{center}
\includegraphics[scale=0.45]{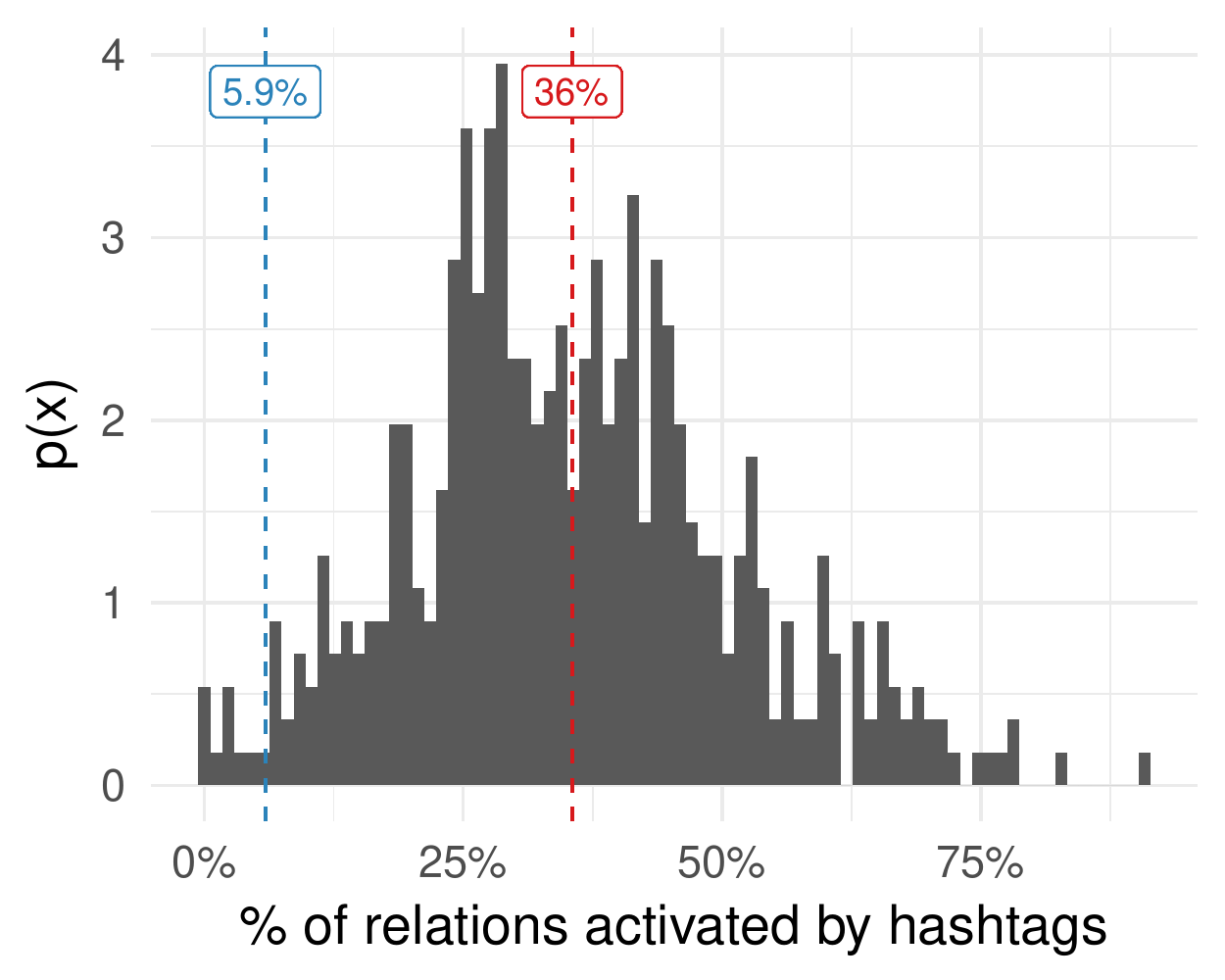}\vspace{-5pt}
\caption{Histogram of the percentage of relations activated by hashtags (in red: mean, in blue: mean for generic Twitter users)}
\label{fig:hist_hashtags}\vspace{-10pt}
\end{center}
\end{figure}

It is reasonable to expect the social tweets generated by journalists to be frequently topic-driven. In order to investigate whether this is the case, in this section we study the relation between alters and hashtags. In Figure~\ref{fig:hist_hashtags}, we show the histogram of the percentage of relationships activated by hashtag per ego. A relationship is considered activated by a hashtag if the first interaction between the ego and the alter contains a hashtag. The average in the journalists dataset is $36\%$, while for generic Twitter users is around $6\%$ and for politicians is around $15\%$~\cite{Arnaboldi2017}. Hence, the Twitter relationships of journalists are very much information-driven.

\begin{figure}[ht]
\begin{center}
\includegraphics[scale=0.45]{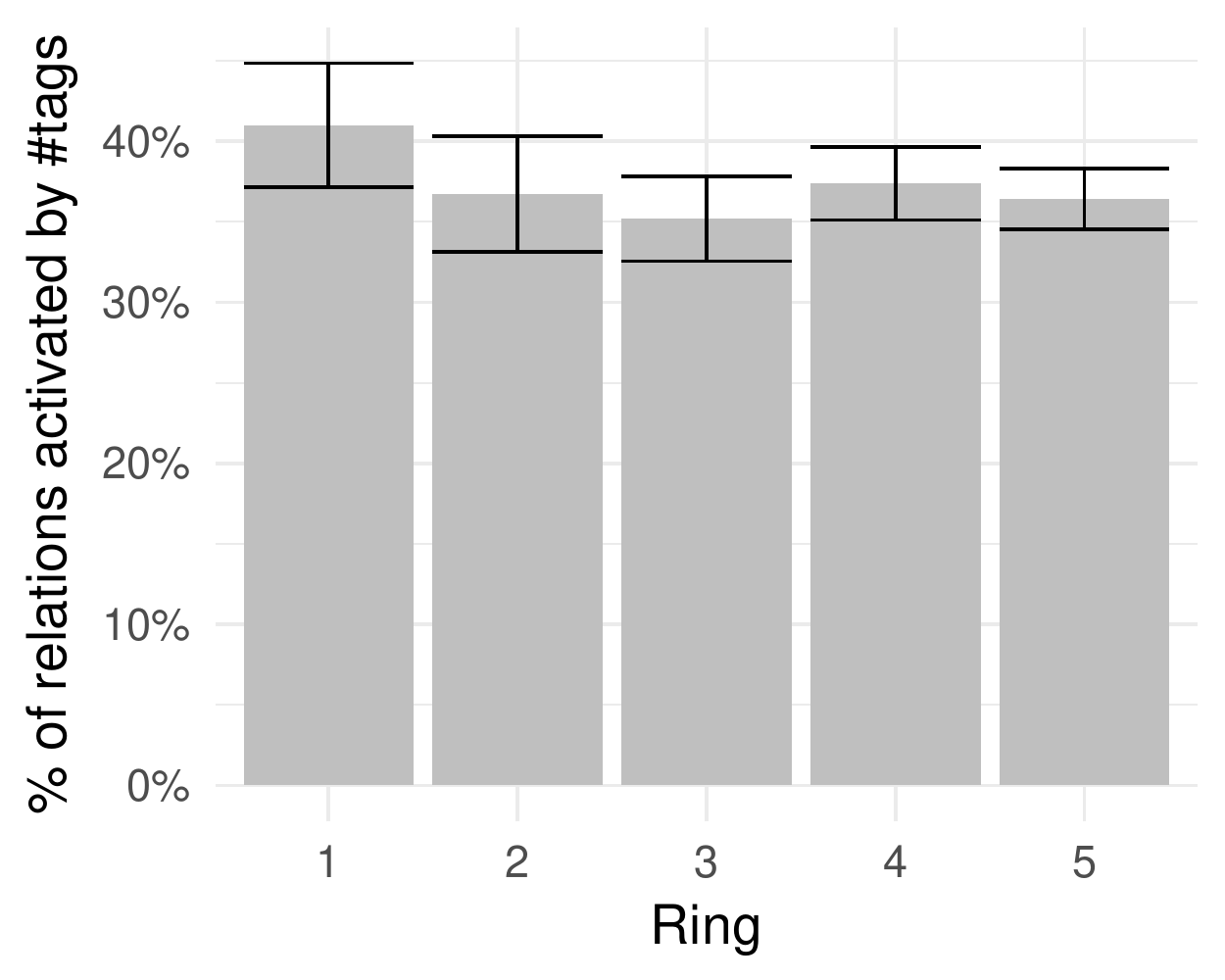}\vspace{-5pt}
\caption{Average percentage of relations activated by hashtags per ring with confidence intervals}
\label{fig:relations_hashtags_rings}\vspace{-10pt}
\end{center}
\end{figure}

\begin{figure}[ht]
\begin{center}
\includegraphics[scale=0.45]{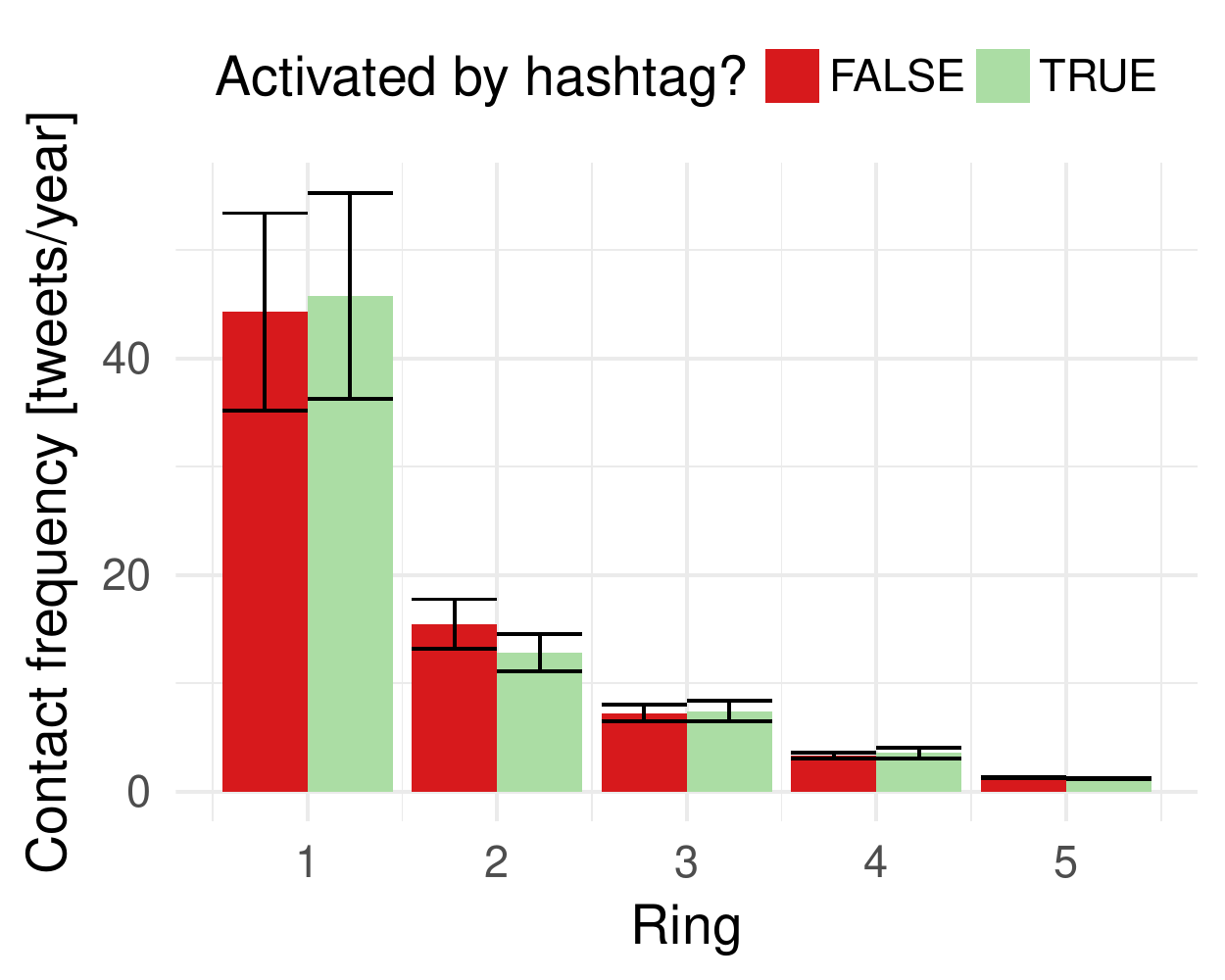}\vspace{-5pt}
\caption{Contact frequency per ring}
\label{fig:contact_freq_rings_hashtags}\vspace{-10pt}
\end{center}
\end{figure}

\begin{figure}[ht]
\begin{center}
\includegraphics[scale=0.45]{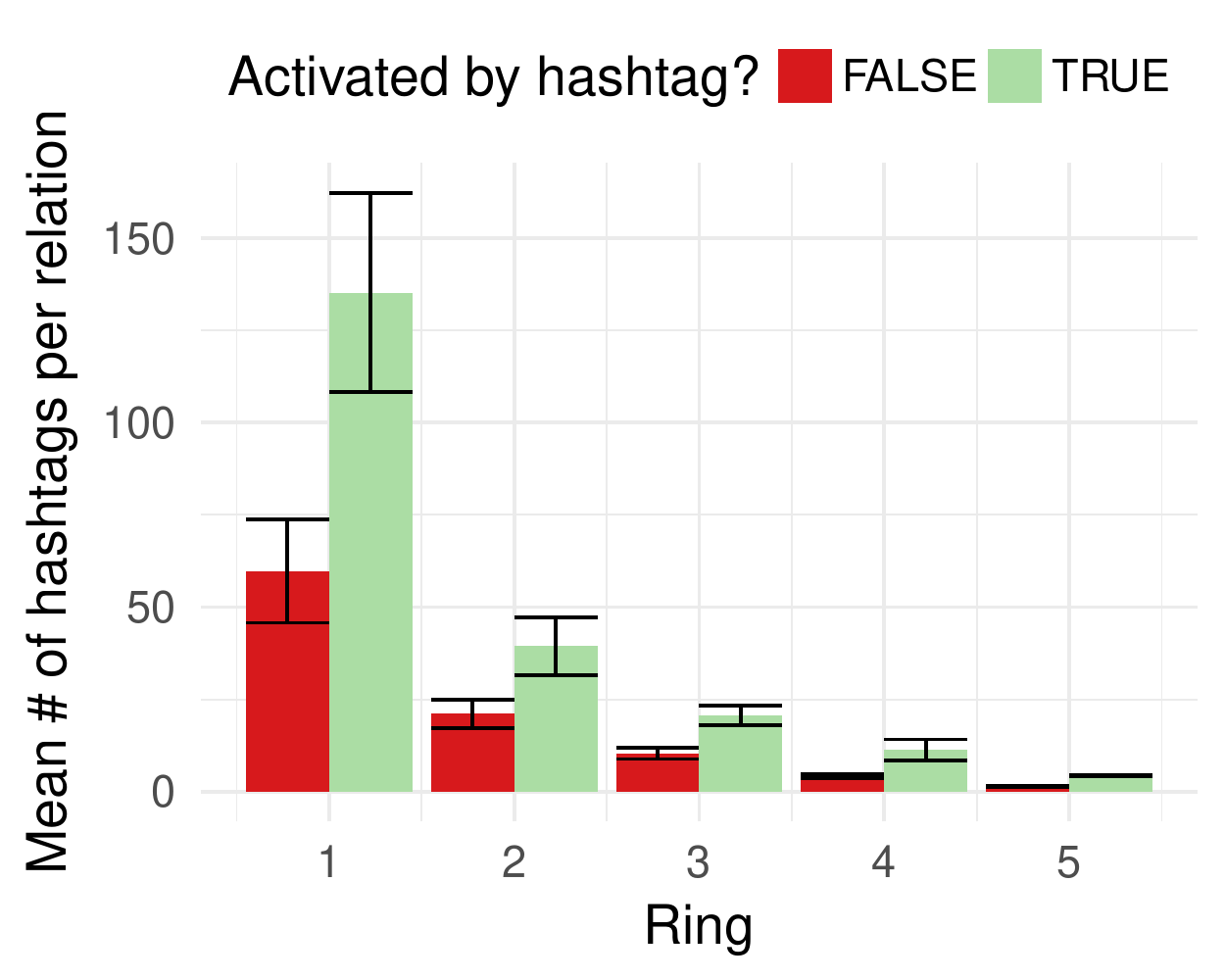}\vspace{-5pt}
\caption{Mean number of hashtags per alter in ring}
\label{fig:mean_hashtags_per_relation}\vspace{-10pt}
\end{center}
\end{figure}

Figure~\ref{fig:relations_hashtags_rings} shows how the activation by hashtag impacts on the different rings. Ring R1 is the most affected, and this was also true for the politicians dataset studied in~\cite{Arnaboldi2017}. However, while the impact of activation-by-hashtag decreases progressively in the outer rings of politicians, it remains stable for journalists. This is again a confirmation that journalists engage on Twitter in an information-driven way, although, as shown by Figures~\ref{fig:jaccard} and~\ref{fig:jump}, the set of alters with which they communicate is very stable, contrarily to politicians and generic users. Figure~\ref{fig:contact_freq_rings_hashtags} shows that there is not a significant difference in contact frequency between alters whose relationship is activated by hashtags and those for which it is not. However, relations that are activated by hashtags tend to use hashtags significantly more often also in future interactions (Figure~\ref{fig:mean_hashtags_per_relation}).

\section{Conclusions}
\label{sec:conclusions}

In this work we have studied the properties of ego networks for a set of the most popular Italian journalists. We have found that the structural patterns of their ego networks mirror those of offline social networks, even more closely than those of generic Twitter users. In particular, both the circle sizes and the circle scaling ratios are very close to the findings from the offline social network literature. Differently from what has been observed for politicians on Twitter, the composition of the circles is quite stable. Journalists stand out in terms of relationships activated by hashtags and number of hashtags per alter, both with respect to generic users and also with respect to politicians. This suggests that journalists' engagement on Twitter is mostly information-driven. 



\begin{acks}
This work was funded by the SoBigData project. The SoBigData project has received funding from the \emph{European Union's Horizon 2020 research and innovation programme} under grant agreement No 654024.
\end{acks}

%% file: osned18-sigconf.bbl
\begin{thebibliography}{18}


\ifx \showCODEN    \undefined \def \showCODEN     #1{\unskip}     \fi
\ifx \showDOI      \undefined \def \showDOI       #1{#1}\fi
\ifx \showISBNx    \undefined \def \showISBNx     #1{\unskip}     \fi
\ifx \showISBNxiii \undefined \def \showISBNxiii  #1{\unskip}     \fi
\ifx \showISSN     \undefined \def \showISSN      #1{\unskip}     \fi
\ifx \showLCCN     \undefined \def \showLCCN      #1{\unskip}     \fi
\ifx \shownote     \undefined \def \shownote      #1{#1}          \fi
\ifx \showarticletitle \undefined \def \showarticletitle #1{#1}   \fi
\ifx \showURL      \undefined \def \showURL       {\relax}        \fi
\providecommand\bibfield[2]{#2}
\providecommand\bibinfo[2]{#2}
\providecommand\natexlab[1]{#1}
\providecommand\showeprint[2][]{arXiv:#2}

\bibitem[\protect\citeauthoryear{Aral and Van~Alstyne}{Aral and
  Van~Alstyne}{2011}]%
        {aral2011diversity}
\bibfield{author}{\bibinfo{person}{Sinan Aral} {and} \bibinfo{person}{Marshall
  Van~Alstyne}.} \bibinfo{year}{2011}\natexlab{}.
\newblock \showarticletitle{The diversity-bandwidth trade-off}.
\newblock \bibinfo{journal}{\emph{Amer. J. Sociology}} \bibinfo{volume}{117},
  \bibinfo{number}{1} (\bibinfo{year}{2011}), \bibinfo{pages}{90--171}.
\newblock


\bibitem[\protect\citeauthoryear{Arnaboldi, Conti, Passarella, and
  Dunbar}{Arnaboldi et~al\mbox{.}}{2013}]%
        {Arnaboldi2013}
\bibfield{author}{\bibinfo{person}{Valerio Arnaboldi}, \bibinfo{person}{Marco
  Conti}, \bibinfo{person}{Andrea Passarella}, {and} \bibinfo{person}{Robin
  Dunbar}.} \bibinfo{year}{2013}\natexlab{}.
\newblock \showarticletitle{{Dynamics of personal social relationships in
  online social networks}}. In \bibinfo{booktitle}{\emph{Proceedings of the
  first ACM conference on Online social networks - COSN '13}}.
  \bibinfo{publisher}{ACM Press}, \bibinfo{address}{New York, New York, USA},
  \bibinfo{pages}{15--26}.
\newblock


\bibitem[\protect\citeauthoryear{Arnaboldi, Passarella, Conti, and
  Dunbar}{Arnaboldi et~al\mbox{.}}{2017}]%
        {Arnaboldi2017}
\bibfield{author}{\bibinfo{person}{Valerio Arnaboldi}, \bibinfo{person}{Andrea
  Passarella}, \bibinfo{person}{Marco Conti}, {and} \bibinfo{person}{Robin
  Dunbar}.} \bibinfo{year}{2017}\natexlab{}.
\newblock \showarticletitle{{Structure of Ego-Alter Relationships of
  Politicians in Twitter}}.
\newblock \bibinfo{journal}{\emph{Journal of Computer-Mediated Communication}}
  \bibinfo{volume}{22}, \bibinfo{number}{5} (\bibinfo{year}{2017}),
  \bibinfo{pages}{231--247}.
\newblock
\showISSN{10836101}


\bibitem[\protect\citeauthoryear{Canter}{Canter}{2015}]%
        {canter2015personalised-tweeting}
\bibfield{author}{\bibinfo{person}{Lily Canter}.}
  \bibinfo{year}{2015}\natexlab{}.
\newblock \showarticletitle{Personalised Tweeting}.
\newblock \bibinfo{journal}{\emph{Digital Journalism}} \bibinfo{volume}{3},
  \bibinfo{number}{6} (\bibinfo{year}{2015}), \bibinfo{pages}{888--907}.
\newblock


\bibitem[\protect\citeauthoryear{Comaniciu and Meer}{Comaniciu and
  Meer}{2002}]%
        {comaniciu2002mean}
\bibfield{author}{\bibinfo{person}{Dorin Comaniciu} {and}
  \bibinfo{person}{Peter Meer}.} \bibinfo{year}{2002}\natexlab{}.
\newblock \showarticletitle{Mean shift: A robust approach toward feature space
  analysis}.
\newblock \bibinfo{journal}{\emph{IEEE Transactions on pattern analysis and
  machine intelligence}} \bibinfo{volume}{24}, \bibinfo{number}{5}
  (\bibinfo{year}{2002}), \bibinfo{pages}{603--619}.
\newblock


\bibitem[\protect\citeauthoryear{Dunbar}{Dunbar}{1998}]%
        {dunbar1998social}
\bibfield{author}{\bibinfo{person}{RI Dunbar}.}
  \bibinfo{year}{1998}\natexlab{}.
\newblock \showarticletitle{The social brain hypothesis}.
\newblock \bibinfo{journal}{\emph{Evolutionary Anthropology}}
  \bibinfo{volume}{9}, \bibinfo{number}{10} (\bibinfo{year}{1998}),
  \bibinfo{pages}{178--190}.
\newblock


\bibitem[\protect\citeauthoryear{Dunbar, Arnaboldi, Conti, and
  Passarella}{Dunbar et~al\mbox{.}}{2015}]%
        {Dunbar2015}
\bibfield{author}{\bibinfo{person}{R.I.M. Dunbar}, \bibinfo{person}{Valerio
  Arnaboldi}, \bibinfo{person}{Marco Conti}, {and} \bibinfo{person}{Andrea
  Passarella}.} \bibinfo{year}{2015}\natexlab{}.
\newblock \showarticletitle{{The structure of online social networks mirrors
  those in the offline world}}.
\newblock \bibinfo{journal}{\emph{Social Networks}}  \bibinfo{volume}{43}
  (\bibinfo{year}{2015}), \bibinfo{pages}{39--47}.
\newblock
\showISSN{03788733}


\bibitem[\protect\citeauthoryear{Everett and Borgatti}{Everett and
  Borgatti}{2005}]%
        {everett2005ego}
\bibfield{author}{\bibinfo{person}{Martin Everett} {and}
  \bibinfo{person}{Stephen~P Borgatti}.} \bibinfo{year}{2005}\natexlab{}.
\newblock \showarticletitle{Ego network betweenness}.
\newblock \bibinfo{journal}{\emph{Social networks}} \bibinfo{volume}{27},
  \bibinfo{number}{1} (\bibinfo{year}{2005}), \bibinfo{pages}{31--38}.
\newblock


\bibitem[\protect\citeauthoryear{Gon{\c{c}}alves, Perra, and
  Vespignani}{Gon{\c{c}}alves et~al\mbox{.}}{2011}]%
        {gonccalves2011modeling}
\bibfield{author}{\bibinfo{person}{Bruno Gon{\c{c}}alves},
  \bibinfo{person}{Nicola Perra}, {and} \bibinfo{person}{Alessandro
  Vespignani}.} \bibinfo{year}{2011}\natexlab{}.
\newblock \showarticletitle{Modeling users' activity on twitter networks:
  Validation of dunbar's number}.
\newblock \bibinfo{journal}{\emph{PloS one}} \bibinfo{volume}{6},
  \bibinfo{number}{8} (\bibinfo{year}{2011}), \bibinfo{pages}{e22656}.
\newblock


\bibitem[\protect\citeauthoryear{Hill and Dunbar}{Hill and Dunbar}{2003}]%
        {hill2003social}
\bibfield{author}{\bibinfo{person}{Russell~A Hill} {and}
  \bibinfo{person}{Robin~IM Dunbar}.} \bibinfo{year}{2003}\natexlab{}.
\newblock \showarticletitle{Social network size in humans}.
\newblock \bibinfo{journal}{\emph{Human nature}} \bibinfo{volume}{14},
  \bibinfo{number}{1} (\bibinfo{year}{2003}), \bibinfo{pages}{53--72}.
\newblock


\bibitem[\protect\citeauthoryear{Java, Song, Finin, and Tseng}{Java
  et~al\mbox{.}}{2007}]%
        {java2007twitter:-understanding}
\bibfield{author}{\bibinfo{person}{Akshay Java}, \bibinfo{person}{Xiaodan
  Song}, \bibinfo{person}{Tim Finin}, {and} \bibinfo{person}{Belle Tseng}.}
  \bibinfo{year}{2007}\natexlab{}.
\newblock \showarticletitle{Why we twitter: understanding microblogging usage
  and communities}. In \bibinfo{booktitle}{\emph{Proceedings of the 9th WebKDD
  and 1st SNA-KDD 2007 workshop on Web mining and social network analysis}}.
  ACM, \bibinfo{pages}{56--65}.
\newblock


\bibitem[\protect\citeauthoryear{Lin, Cook, and Burt}{Lin
  et~al\mbox{.}}{2001}]%
        {lin2001social}
\bibfield{author}{\bibinfo{person}{Nan Lin}, \bibinfo{person}{Karen~S Cook},
  {and} \bibinfo{person}{Ronald~S Burt}.} \bibinfo{year}{2001}\natexlab{}.
\newblock \bibinfo{booktitle}{\emph{Social capital: Theory and research}}.
\newblock \bibinfo{publisher}{Transaction Publishers}.
\newblock


\bibitem[\protect\citeauthoryear{McCarty}{McCarty}{2002}]%
        {mccarty2002structure}
\bibfield{author}{\bibinfo{person}{Christopher McCarty}.}
  \bibinfo{year}{2002}\natexlab{}.
\newblock \showarticletitle{Structure in personal networks}.
\newblock \bibinfo{journal}{\emph{Journal of social structure}}
  \bibinfo{volume}{3}, \bibinfo{number}{1} (\bibinfo{year}{2002}),
  \bibinfo{pages}{20}.
\newblock


\bibitem[\protect\citeauthoryear{Quercia, Capra, and Crowcroft}{Quercia
  et~al\mbox{.}}{2012}]%
        {quercia2012social}
\bibfield{author}{\bibinfo{person}{Daniele Quercia}, \bibinfo{person}{Licia
  Capra}, {and} \bibinfo{person}{Jon Crowcroft}.}
  \bibinfo{year}{2012}\natexlab{}.
\newblock \showarticletitle{The Social World of Twitter: Topics, Geography, and
  Emotions.}
\newblock \bibinfo{journal}{\emph{ICWSM}}  \bibinfo{volume}{12}
  (\bibinfo{year}{2012}), \bibinfo{pages}{298--305}.
\newblock


\bibitem[\protect\citeauthoryear{Russell, Hendricks, Choi, and
  Stephens}{Russell et~al\mbox{.}}{2015}]%
        {russell2015sets}
\bibfield{author}{\bibinfo{person}{Frank~Michael Russell},
  \bibinfo{person}{Marina~A Hendricks}, \bibinfo{person}{Heesook Choi}, {and}
  \bibinfo{person}{Elizabeth~Conner Stephens}.}
  \bibinfo{year}{2015}\natexlab{}.
\newblock \showarticletitle{Who Sets the News Agenda on Twitter? Journalists'
  posts during the 2013 US government shutdown}.
\newblock \bibinfo{journal}{\emph{Digital Journalism}} \bibinfo{volume}{3},
  \bibinfo{number}{6} (\bibinfo{year}{2015}), \bibinfo{pages}{925--943}.
\newblock


\bibitem[\protect\citeauthoryear{Sutcliffe, Dunbar, Binder, and
  Arrow}{Sutcliffe et~al\mbox{.}}{2012}]%
        {sutcliffe2012relationships}
\bibfield{author}{\bibinfo{person}{Alistair Sutcliffe}, \bibinfo{person}{Robin
  Dunbar}, \bibinfo{person}{Jens Binder}, {and} \bibinfo{person}{Holly Arrow}.}
  \bibinfo{year}{2012}\natexlab{}.
\newblock \showarticletitle{Relationships and the social brain: integrating
  psychological and evolutionary perspectives}.
\newblock \bibinfo{journal}{\emph{British journal of psychology}}
  \bibinfo{volume}{103}, \bibinfo{number}{2} (\bibinfo{year}{2012}),
  \bibinfo{pages}{149--168}.
\newblock


\bibitem[\protect\citeauthoryear{Tumasjan, Sprenger, Sandner, and
  Welpe}{Tumasjan et~al\mbox{.}}{2010}]%
        {tumasjan2010predicting}
\bibfield{author}{\bibinfo{person}{Andranik Tumasjan},
  \bibinfo{person}{Timm~Oliver Sprenger}, \bibinfo{person}{Philipp~G Sandner},
  {and} \bibinfo{person}{Isabell~M Welpe}.} \bibinfo{year}{2010}\natexlab{}.
\newblock \showarticletitle{Predicting elections with twitter: What 140
  characters reveal about political sentiment.}
\newblock \bibinfo{journal}{\emph{Icwsm}} \bibinfo{volume}{10},
  \bibinfo{number}{1} (\bibinfo{year}{2010}), \bibinfo{pages}{178--185}.
\newblock


\bibitem[\protect\citeauthoryear{Zhou, Sornette, Hill, and Dunbar}{Zhou
  et~al\mbox{.}}{2005}]%
        {Zhou2005}
\bibfield{author}{\bibinfo{person}{W-X Zhou}, \bibinfo{person}{D Sornette},
  \bibinfo{person}{R~a Hill}, {and} \bibinfo{person}{R~I~M Dunbar}.}
  \bibinfo{year}{2005}\natexlab{}.
\newblock \showarticletitle{{Discrete hierarchical organization of social group
  sizes.}}
\newblock \bibinfo{journal}{\emph{Proceedings. Biological sciences / The Royal
  Society}} \bibinfo{volume}{272}, \bibinfo{number}{1561}
  (\bibinfo{year}{2005}), \bibinfo{pages}{439--444}.
\newblock


\end{thebibliography}
